# Characteristics of a Plasma Source with adjustable multi-pole line cusp geometry

Meenakshee Sharma[1,2], A. D. Patel[1], N. Ramasubramanian[1,2], Y. C. Saxena[1], P. K. Chattopadhyaya[1,2], and R. Ganesh[1,2]

[1]Institute for Plasma Research, Bhat Gandhinagar-382428

[2]Homi Bhabha National Institute, Anushakti Nagar, Mumbai-400094

**Abstract**

Two magnetic configurations of Multi-cusp Plasma Device (MPD) have been explored to obtain high quiescence level, large uniform plasma region with nearly flat mean density and temperature profiles. In particular, properties of plasma in a six-pole six magnet (SPSM) and twelve pole six magnets (TPSM) cusp configurations are rigorously compared and reported here. It is found that more uniform plasma with nearly flat profiles is found in TPSM along with increased quiescence level. Findings are verified across various magnetic field strengths for both configurations.

## I. Introduction

The production of large volume, uniform, and quiescent plasma has been a subject of research since many decades because of its various applications, such as material plasma processing [1–8], plasma wall interaction for material selection towards plasma facing component in ITER [9]. It is also desirable in thermonuclear fusion due to its favorable magnetohydrodynamic stability [10–13]. Limpaecher and MacKenzie [14] first showed that the multi-pole cusp magnetic field could be used to produce such plasmas with enhanced densities. Such magnetic field enhances the plasma density ∼100 times and makes the plasma more quiescent and uniform. The energetic ionizing electrons are confined in the plasma by the magnetic fields at the walls resulting in a high ionization efficiency [15]. Because of the quiescent nature, the plasma confined by multipole magnetic field has also been found to be useful for studies of fundamental plasma phenomenon, such as the study of excitation and propagation of various waves in plasma, wave particle trapping and un-trapping, Landau damping of waves [16,17], phase mixing [18,19] and wave breaking [20–22].

The Multi-pole line cusp Plasma Device (MPD) [23] has adopted this multi-pole cusp magnetic field geometry for plasma confinement. The usage of electromagnets in MPD facilitates the confinement of plasma with various cusp geometries and a wide range of magnetic field strengths. The characteristics of the filamentary produced argon plasma in MPD have been studied and reported before [23,24]. Drift waves [25] have been observed in the edge region of the Six Pole Six Magnet (SPSM) configurations in which the wave vector changing its direction has also been seen for the first time. Since these drift wave fluctuations cannot reach the central regions because of the good curvature physics, the plasma in the central region of the MPD is found to be very quiescent. This quiescent and uniform plasma volume can be used to study fundamental plasma phenomena as well as for plasma applications mentioned above. Since the cusp fields are produced by electromagnets, the number of pole cusps can be changed by changing the direction of the current in the coils. Thus by passing the current in the same direction in all the six magnets, a Twelve Pole Six Magnet (TPSM) line cusp configuration can be created in the same volume. Because of more poles, the volume with small magnetic field will increase and can thus get a larger volume of uniform and quiescent plasma. This larger uniform and quiescence plasma volume allows to observe the plasma interaction and response to any external perturbation to excite the various waves. This article reports the detailed plasma characteristics in the Twelve Pole Six Magnet (TPSM) configuration as compared to the Six Pole Six Magnet (SPSM) configurations while bringing out the advantageous property of the former. The rest of the paper is organized as follows, the schematic of the device and diagnostics are described in Section II, and Section III contains the experimental observations and discussion followed by a summary in section IV.

## II. Experimental Setup and Diagnostics

### i. Plasma production

The experiments reported here are performed in a linear cylindrical device MPD, which has been described before in previous studies [23,24]. The schematic of the MPD, its diagnostics, and coordinate system is shown in Figure 1. The device has a cylindrical chamber of 150 cm length, 40 cm diameter and 0.6 cm thick wall made of stainless steel. The system is pumped down to $2 \times 10^{-6}$ mbar base pressure and the operating pressure of filled Argon gas varies from $5 \times 10^{-5}$ mbar to $3 \times 10^{-3}$ mbar. A rectangular filamentary array of $8 \times 8$ cm², centered at R=0 cm, is used as a cathode to produce the plasma.

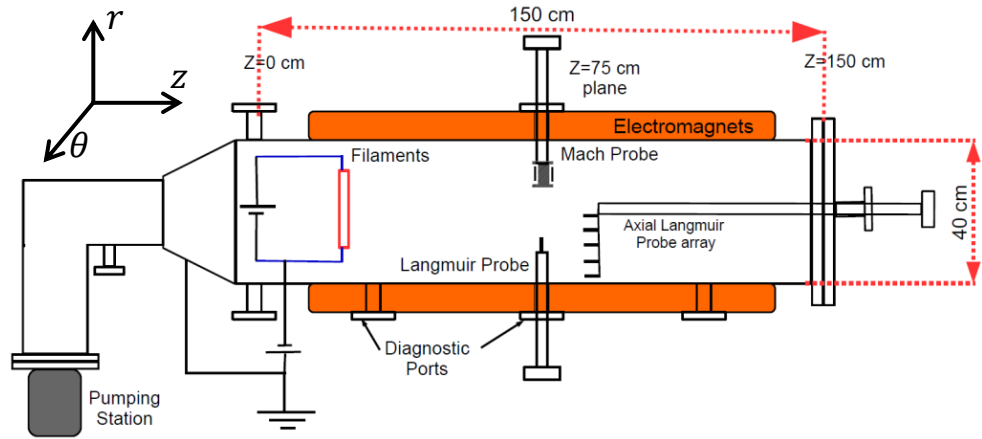

*Figure 1: Schematic of the side view of the MPD device, electromagnet, and its diagnostics, inset of the figure shows the opted coordinate system of the device. All the measurements have been performed on $z = 75$ cm.*

The filaments are heated for thermionic emission of electrons using a 15V, 500A floating power supply biased at $-50$V with respect to an anode (wall of the cylindrical chamber) using 128V, 25A power supply.

## ii. Magnetic field configuration

Six electromagnets with 120 cm long rectangular bars of vacoflux embedded in copper coils are installed on the periphery of the cylindrical chamber with $60^o$ azimuthal spacing. More details about these magnets are reported here [16]. The use of electromagnets in MPD, gives the flexibility to work with different multi-cusp geometries by changing the current directions in the said coils, while the field strength can be varied by changing the current values. These flexibilities give the freedom to study changes in the plasma parameters with respect to different field values and configurations. The direction of current in these electromagnets decides the North (N) and South (S) poles of each. In this paper, the results from two different configurations are presented.

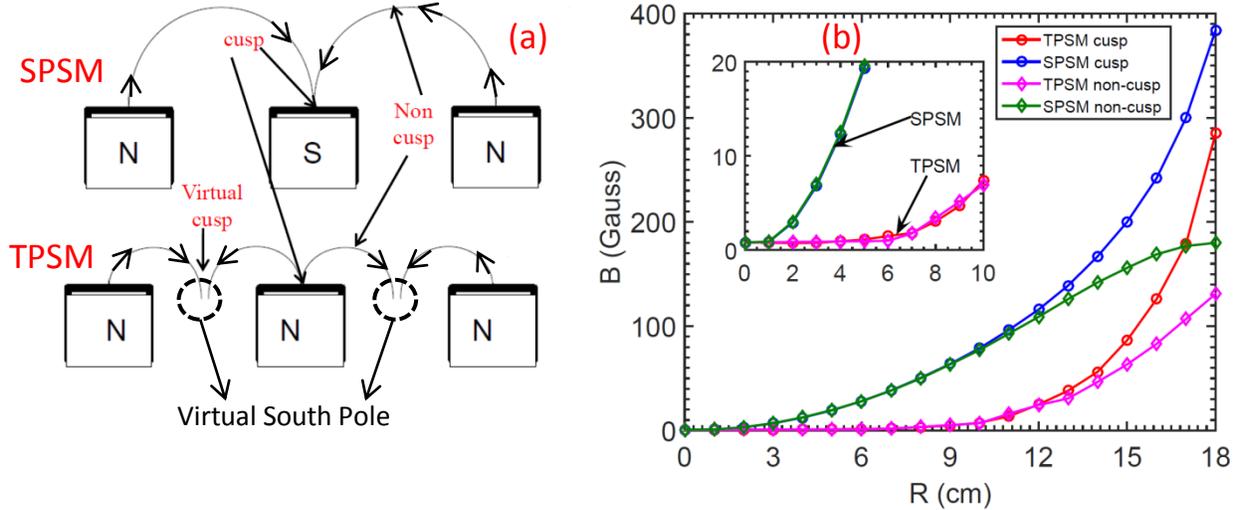

*Figure 2: (a) The schematic of SPSM and TPSM. The non-cusp (the regime from the centre of the device to exactly middle of the two pole surface), and cusp (the region from center of the device to pole of the magnets in TPSM the regime from centre of the device to virtual pole is also cusp regime). (b) Radial variation of magnetic field strength for six pole six magnet (SPSM) and twelve pole six magnet (TPSM) cusp configuration.*

The first one is called Six Poles with Six Magnets (SPSM), in which the current in alternate coils will have opposite directions thus each magnet acts a pole cusp for totally six poles of three dipoles. A part of this configuration is shown in the first line of figure 2(a). The other one is called Twelve Poles with Six Magnets (TPSM) in which the current in all six magnets are in the same direction. In this configuration, while all the magnets will produce one type of poles, say north-pole, six virtual complementary poles (south) will be created in between the magnets, thus having totally 12 poles of six dipoles. A part of this configuration is shown in the bottom line of the same figure 2(a). The radial variation of net field values measured using Hall probes as well as simulated values using FEMM [26] are shown in figure 2(b) for both SPSM and TPSM.

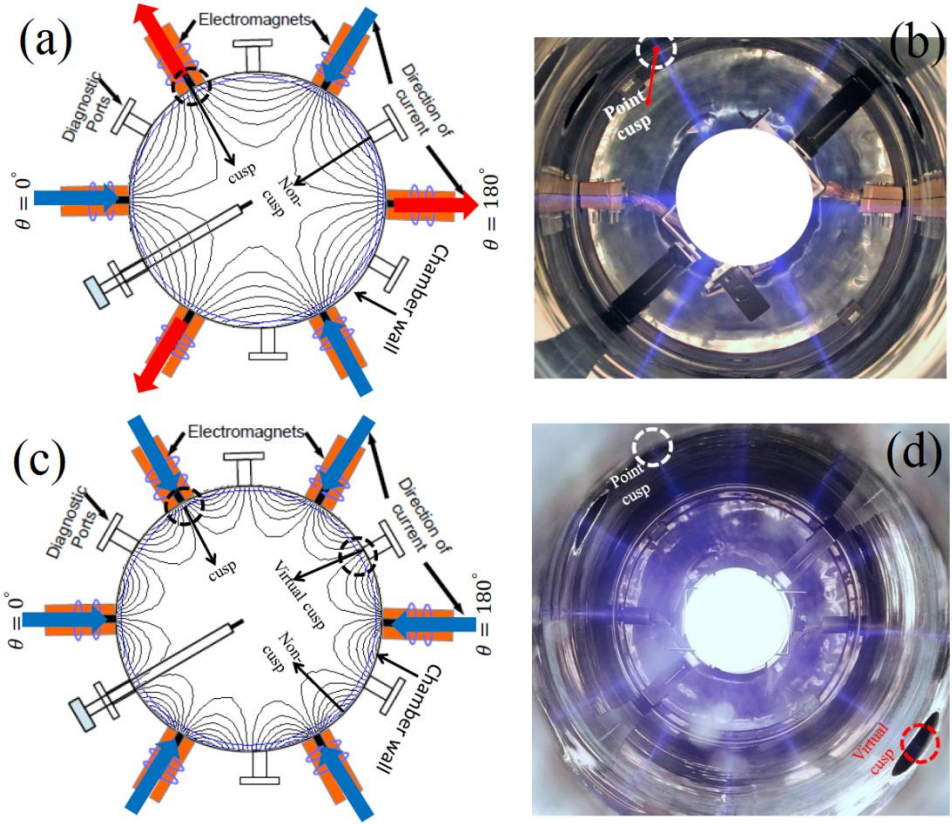

*Figure 3: (a) and (c) shows the arrangement of electromagnets over the chamber, diagnostic ports, Langmuir probe for diagnostic and magnetic field lines simulated using FEMM in SPSM and TPSM configuration respectively. Arrow in figures shows the current direction in electromagnets. Figure (b) and (d) shows the pictures of plasma confined in SPSM and TPSM observed through the viewport from one end of the device, the center region of bright glow of filaments has been shadowed to capture the feeble light from wings or the cusp regions.*

It can be seen that the net field values for TPSM are less along more radial extent compared to the SPSM which means the radial extent of field-free regions is more in TPSM as expected. Figures 3(a) and 3(c) show the cross-sectional field lines simulated using FEMM for SPSM and TPSM respectively. The arrows marked in these figures show the measurement paths for plasma parameters along the cusp region and the non-cusp regions. Figure 2(b) also shows the radial variation of the field all along those arrows through a cusp region (parallel to the field) and a non-cusp (across the field lines) regions. The photos of the argon plasma in those two configurations (viz. SPSM and TPSM) are shown in figure 3 which are taken from the machine end opposite to the source. In figure 3(b) and 3(d) we have to write that the bright glow of filaments has been shadowed to capture the feeble light from wings or the cusp regions. Six wings structure representing six cusps in the SPSM (figure 3(d)) and twelve of the same (figure 3(b)) for TPSM can be seen in the respective figures. It is because the electrons travel along the magnetic field lines thus ionizing and exciting line radiations on the way, till the field lines touch the walls of the vacuum chamber. It can also be seen from the photographs that the radial extent of visible light, hence plasma, is more in the TPSM compared to the SPSM. This is because of the more field free region in the TPSM.

iii. **Diagnostic technique**

The characterization of the plasma in both configurations is done using Langmuir probes and Mach probes. The Langmuir probes are of 1 mm diameter and 5 mm length and are used to deduce the plasma parameters like electron temperature ($T_e$), plasma density ($n$), plasma potential ($V_p$), floating potential ($V_f$) and fluctuations. The details about the probe measurements and the corrections applied to deduce the respective values are given in the references [16, 20, 21].

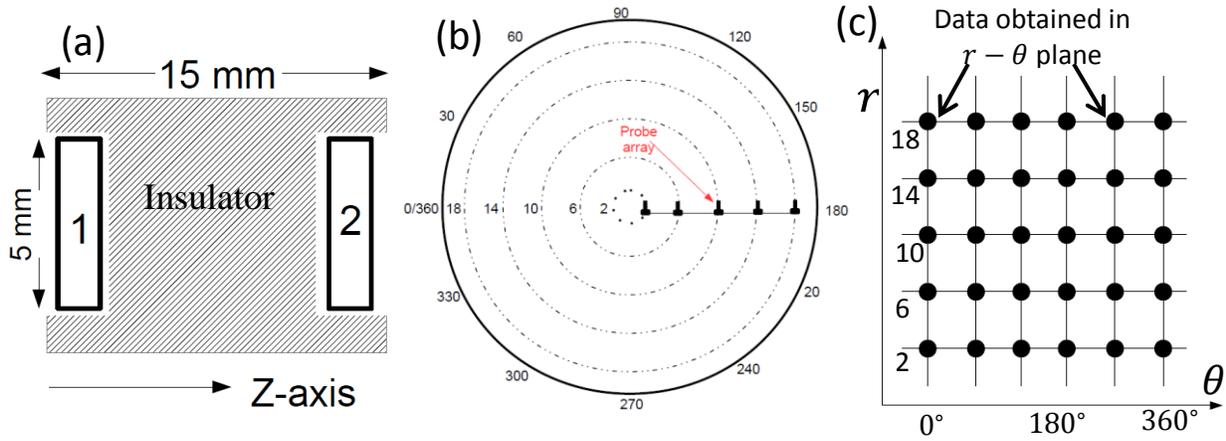

*Figure 4: Schematic of (a) Mach probe, (b) five radial probe array integrated with a rotatable axial shaft, (c) 2-d measurement in $r - \theta$ plane.*

The Mach probe used here is a pair of two disk probes of 5 mm diameter each, separated by a ceramic insulator and is mounted so as to measure the flow in the axial direction (z) as shown in figure 4(a). This probe is constructed and operated according to the details given in the references [27,28]. Probe 1 is upstream collector facing the filaments and probe 2 is a downstream collector in the direction of the z-axis respectively. The Mach probe is used to determine the radial profile of the axial plasma flow. Both plates are biased to negative potential to measure ion saturation current which will be further used to calculate the Mach number ($u_n$). The plasma flow ($V_d$) can be calculated using Mach number ($u_n$) and ion-acoustic velocity ($C_s = \sqrt{KT_e/m_i}$) measured by Langmuir probe, $V_d = u_n C_s$. In Mach probe, area of the collectors should be same. For reliability and authenticity of the results, the probe is rotated by 180 degrees and ion saturation current is measured. The ratio of ion saturation current for both cases, the previous arrangement and after the 180 degree rotation is compared and found to be the same which implies the equal probe area. While the radial variation of plasma parameters along the non-cusp region of SPSM using the radial port, the measurement along the non-cusp region of TPSM is not possible. Hence a linear Langmuir probe array is constructed to measure the azimuthal variation of ion saturation current which covers the cusp as well as non-cusp regions of both TPSM and SPSM. This array consists of 5 probes placed at five radial locations with an interval of 4 cm i.e., at R=2, 6, 10, 14, and 18 cm, as shown in figure 4(b). This array is mounted on a rotatable axial shaft and the shaft itself is fitted to the endplate of the vacuum chamber such that the shaft is at the center of the device at R=0. This arrangement gives a 2-d measurement on the r-θ plane as shown in figure 4(c). The details of the experimental results obtained using the above mentioned diagnostics are presented and discussed in the following section III.

## III. Experimental results and discussion

Though the experiments were carried out for many pressure values and many discharge voltages, for the comparison of SPSM and TPSM configurations, the data from experiments with $2 \times 10^{-4}$ mbar Argon pressure and -50 V discharge voltage are only considered here for this report. The radial variation of mean plasma parameters are measured at $z = 75$ cm, on the vertical midplane of the device.

### A. Characterization of Mean Plasma Parameters

#### i. *Variation of Plasma Parameters with Cusp Magnetic Field Strength at the center*

It is well documented that the cusp magnetic field traps the primary electrons by the mirror effect [29–33]. The number of electrons trapped and their maximum energy for given a discharge conditions depends upon the field strength at the pole cusp which depends on the magnitude of current. If the current passed in both SPSM and TPSM configurations are same in magnitude, the pole strength in each cusp will be same in both configurations. The observed variation of plasma parameters with increasing cusp magnetic field strength measured at R=0 cm for both cusp configuration are shown in figure 5.

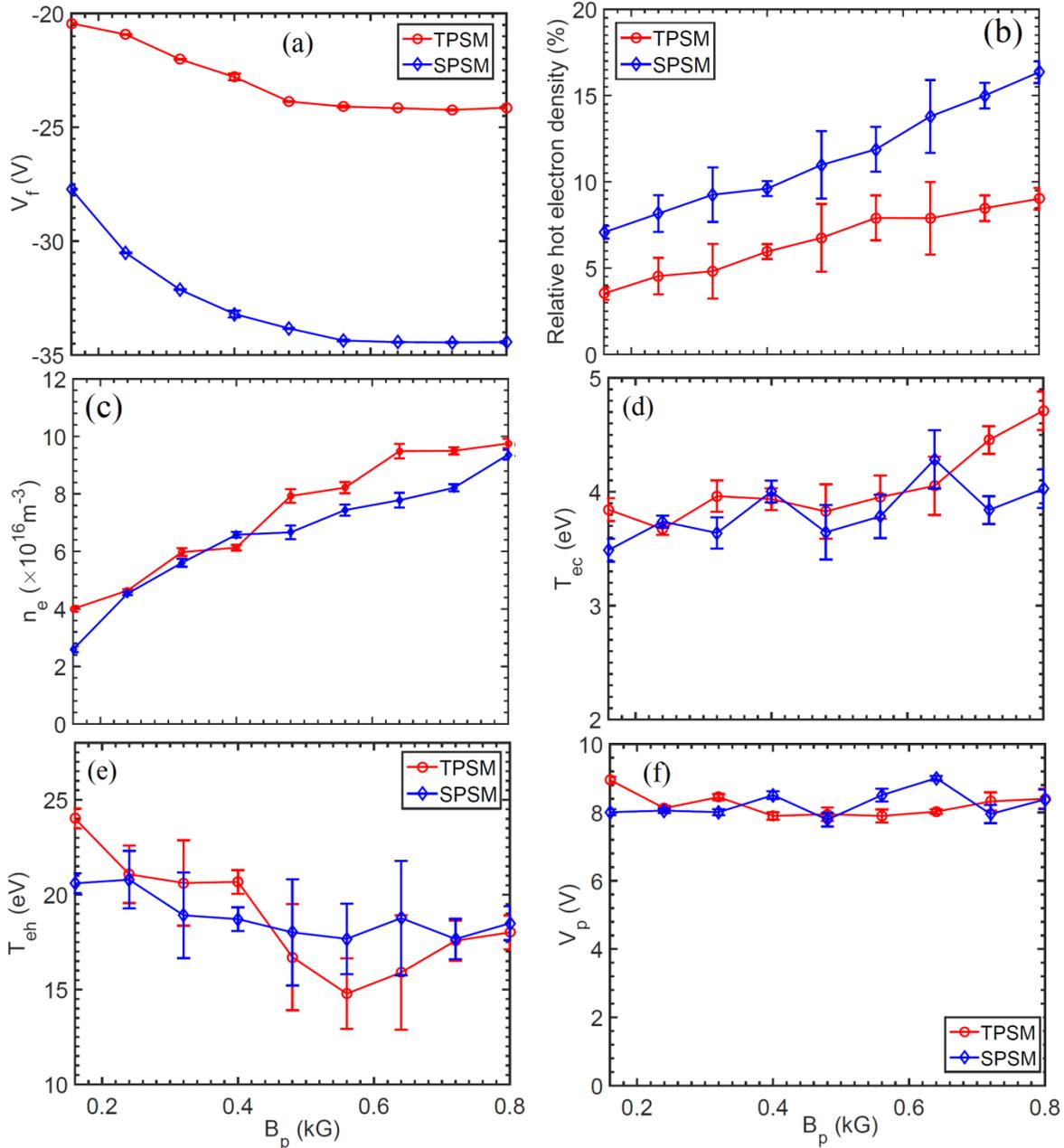

Figure 5: Variation of plasma parameters with various cusp magnetic field strength (a) floating potential, $V_f$, (b) hot electron density, $n_{eh}$, (c) plasma density, $n_e$, (d) temperature of bulk electron, $T_{ec}$, (e) temperature of hot electron, $T_{eh}$, and (f) plasma potential, $V_p$ measured at R=0 cm, for SPSM and TPSM cusp configuration at $2 \times 10^{-4}$ mbar pressure and -50 V discharge voltage.

The line with circle and line with diamond shape marker shows plasma parameters for TPSM and SPSM respectively. Figure 5(a) shows the variation of floating potential ($V_f$), the value of floating potential is relatively more negative in SPSM compare to TPSM for all magnetic field values. It indicates higher confinement of hot electrons in SPSM relative to TPSM. This can also be confirmed in figure 5(b), which shows the relative hot electron population for both cusp configurations. The hot electrons density relative to plasma density, $(n_{eh}/n_e) \times 100$), varies from $7\% - 15\%$ in SPSM and $0.7\% - 8\%$ in TPSM; nearly twice in SPSM. It indicates that SPSM can confine the same amount of hot electrons with fewer magnetic field. As reported earlier [24] as the cusp field strength increases, the leak width [34,35] of plasma (plasma loss to the chamber wall while following the field lines) decreases. As a consequence of this, the population of the hot electron increases which further leads to an increase in the magnitude of plasma density. Plasma density in both cusps increases with an increase in the cusp magnetic field strength as shown in figure 5(c). It is clearly observed from the figure, that the values of plasma density are nearly comparable in both cusp configurations for equal cusp magnetic field strength. The variation of plasma bulk electron temperature, $T_{ec}$ with increasing cusp, magnetic field strength is shown in figure 5(d). The bulk electron temperature does not vary much with respect to cusp magnetic field strength and it is nearly same for both TPSM and SPSM. It suggests that the physical mechanism that governs the electron temperature in plasma is likely to be independent of the magnetic field. The variation of hot electron temperature, $T_{eh}$ with increasing cusp magnetic field strength is shown in figure 5(e) it does not vary much with respect to cusp magnetic field strength and is comparable for both TPSM and SPSM. The energy of hot electrons depends upon the discharge voltage and neutral density. Both the parameters are kept constant at $-50$ V and $2 \times 10^{-4}$ mbar for both types of configurations. Therefore the temperature of hot electrons does not vary with the cusp magnetic field. The plasma potential $V_p$ is also nearly constant with the variation of cusp magnetic field strength in both the configurations and observed to have similar values, as shown in figure 5(f).

From figure 5, it can be observed that confinement of hot electrons is more in SPSM compared to TPSM. The various other plasma parameters such as plasma density, electron temperature, and plasma potential show comparable values for both cusp configurations and follows a similar trend with increasing cusp magnetic field strength.

## ii. Azimuthal Variation of Ion Saturation Current

In SPSM, there are 6- cusp regions and 6 non-cusp regions, and the data is obtained at 12 azimuthal locations. In TPSM, there are 6- cusp, 6-virtual cusp and 12-non-cusp regions and the data are obtained at 24 azimuthal points. The probe array is used to measure the azimuthal variation of ion saturation current, at all cusp and non-cusp regions for five radial locations, at Z=75cm. The measured profile of azimuthal variation of ion saturation current is shown in figure 6(a) and 6(b) for TPSM and SPSM respectively. The x-axis of both figures represents azimuthal locations in $\theta$ coordinates. It can be observed from figure 6(a) that in TPSM at R=2, 6, and 10 cm, values of ion saturation current do not vary much azimuthally, traversing from cusp to non-cusp regions. At R=14cm and 18cm, it can be observed that the value of ion saturation current is higher in the cusp region (0.18 mA) but it falls down in non-cusp region (0.08 mA) while the values are changing for cusp and virtual cusp regions. Figure 6(b) shows the azimuthal variation of ion saturation in, SPSM. Beyond R=2 cm, the values changes noticeable at the cusp and non-cusp regions.

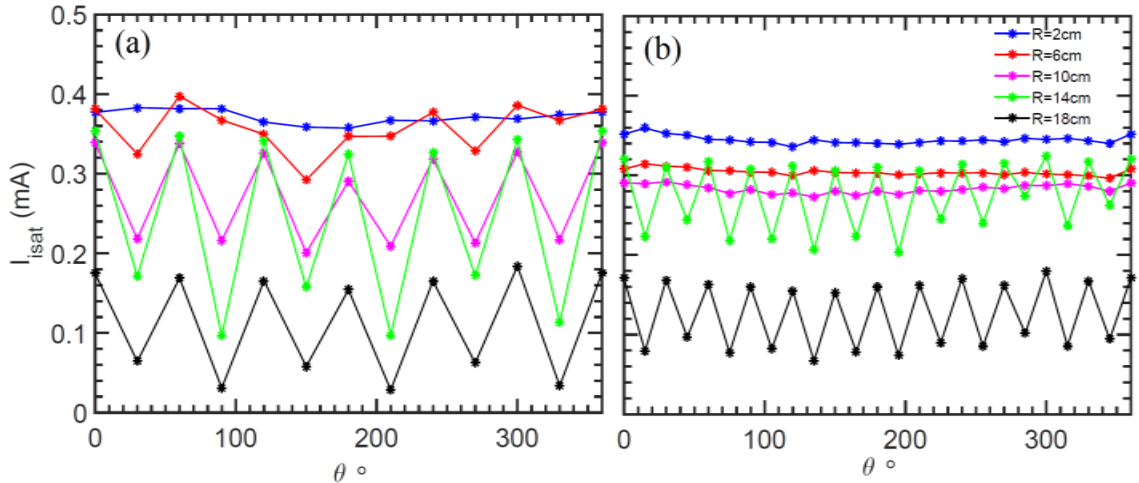

Figure 6: Azimuthal variation of ion saturation current (a) in SPSM, and (b) TPSM for five radial location R=2, 6, 10, 14, and 18 cm, measure at Z=75 cm, midplane of the device for $2 \times 10^{-4}$ mbar, -50 V discharge voltage and $\boldsymbol{B_p} = \boldsymbol{400}$ Gauss.

From figure 6, it can also be observed that the decrease in ion saturation current from cusp to non-cusp is observed to be larger in SPSM (~0.18 to ~0.02 mA) compared to TPSM (~0.18 to ~0.08 mA). This shows the volume of uniform plasma in the TPSM is larger than the volume in SPSM which matches with the field free region volume.

### *iii.* *Radial Variation of Plasma Parameters*

The radial measurement is only possible in between two consecutive magnets, due to the limitation of the device. The location of the radial measurements is shown in figure 3(a) and 3(c), this location is a virtual cusp region in TPSM and non-cusp region in SPSM. The non-cusp region of TPSM and cusp region of SPSM are not accessible radially; therefore, the axial probe array is used to measure plasma parameters in these regions but only at five radial locations.

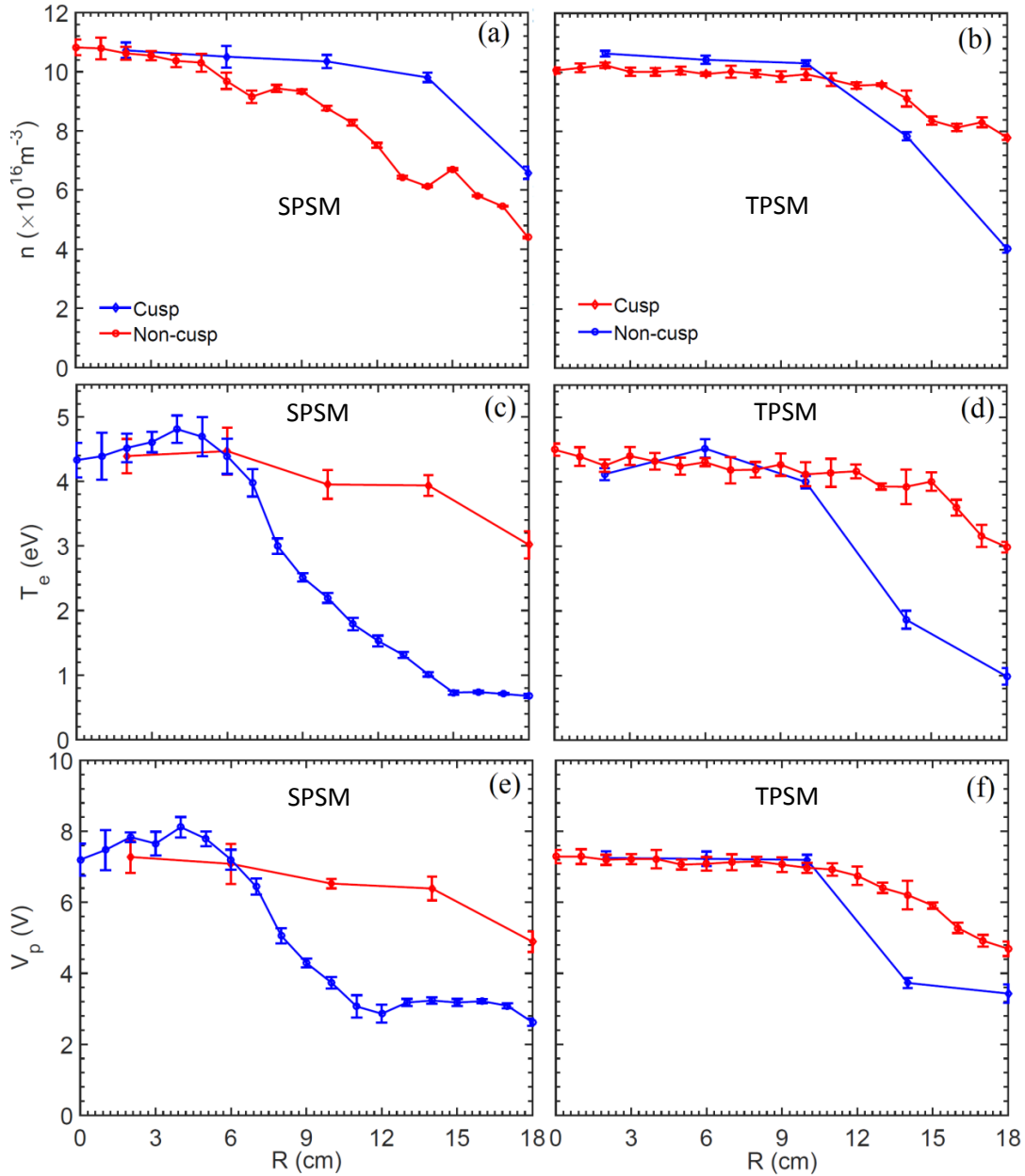

*Figure 7: Radial variation of (a), (b) plasma density, n, (c), (d) bulk plasma electron temperature, $T_e$, and (e), (f) plasma potential, $V_p$, for SPSM and TPSM respectively. $B_p = 800$ Gauss, $2 \times 10^{-4}$ mbar neutral pressure, and $-50$ V discharge voltage.*

The radial profiles of plasma parameters for both types of cusp configurations are shown in figure 7. The radial profiles of plasma density are shown in figure 7(a) and 7(b). In SPSM, $< 5\%$ variation in density is observed upto $R \sim 4$ cm, as shown in figure 7(a). The plasma density is nearly constant and varies $< 5\%$ upto $R \sim 10$ cm for TPSM which corresponds to the respective field free region for the same configuration. Beyond this spatial location, the plasma density gradually decreases towards the chamber wall as shown in figure 7(b). The density has a sharper gradient for non-cusp region of SPSM compared to TPSM. Figure 7(c) shows the radial variation of electron temperature in SPSM configuration. In cusp region, the temperature is nearly constant and shows a shallower gradient after $R \sim 10$ cm. On the other hand, the temperature has a sharp gradient around $R \sim 4$ cm in non-cusp region of SPSM and becomes flat near to $R \sim 14 - 18$ cm. Figure 7(d) shows the radial variation of electron temperature for TPSM, it suggests that the bulk electron temperature is nearly constant upto $R \sim 10$ cm and then it has a shallower gradient around R=12-14 cm in cusp region and fast fall is observed in non-cusp region. The plasma potential shown in figure 7(e) and 7(f) shows a similar trend as electron temperature in both cusp configurations. It is uniform upto $R \sim 10$ cm in cusp and non-cusp region of TPSM and decreases gradually in cusp region compare to non-cusp. Plasma potential is used to estimate the electric field in plasma. The radial profile of plasma potential suggests, that radial electric field is either not present or it is very feeble upto $R \sim 10$ cm in TPSM followed by a very weak gradient outboard for other radial locations. The non-cusp region of SPSM has a sharp gradient in electric field from $R \sim 4 - 10$ cm and it is negligible afterwards. From figure 7, larger radial extent of uniformity in all plasma parameters is observed in TPSM ($R \sim 10$ cm) in comparison to SPSM ($R \sim 4$ cm).

## B. Characterization of Plasma Fluctuations

### i. Quiescence Level

The level of fluctuation in plasma density is measured for both configurations and its radial variation is shown in figure 8. The figure 8 shows the relatively lower ($< 0.1\%$) values of level of fluctuation in density for both configuration upto the radial extent of $\sim 4 - 5$ cm. The level of fluctuation in SPSM increases radially outward from $R \sim 4 - 5$ cm, rising sharply near to wall of the device. The level of fluctuation in TPSM has a constant low (<0.1%) values upto $R \sim 14$ cm rising to ~0.8% towards wall of the device.

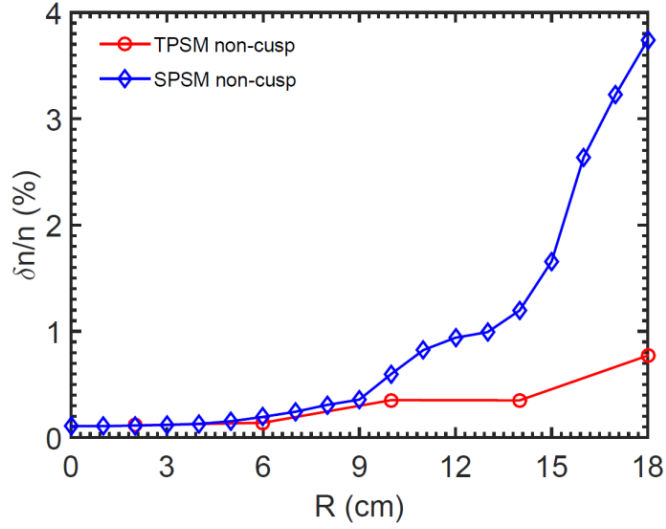

*Figure 8: Radial variation of level of fluctuation in plasma density measured for pole cusp magnetic field $B_p = 800$ Gauss, $2 \times 10^{-4}$ mbar neutral pressure and $-50$ V discharge voltage. The line with circle is for TPSM and line with diamond shape marker is for SPSM.*

The radial extent where the plasma density variation is uniform and <5%, the density fluctuation is $< 0.1\%$ and uniform. The value of level of fluctuation near to chamber wall at $R = 18$ cm is less in TPSM ($\sim 0.8\%$) compare to SPSM ($\sim 4\%$). A detailed study of density fluctuation in SPSM near to wall is reported earlier [25], in which the presence of drift wave due to the gradient in plasma density is observed in non-cusp regime. The same experiments are carried out in non-cusp regime of SPSM and TPSM to determine the nature of fluctuations and to explore the presence of the drift wave mode.

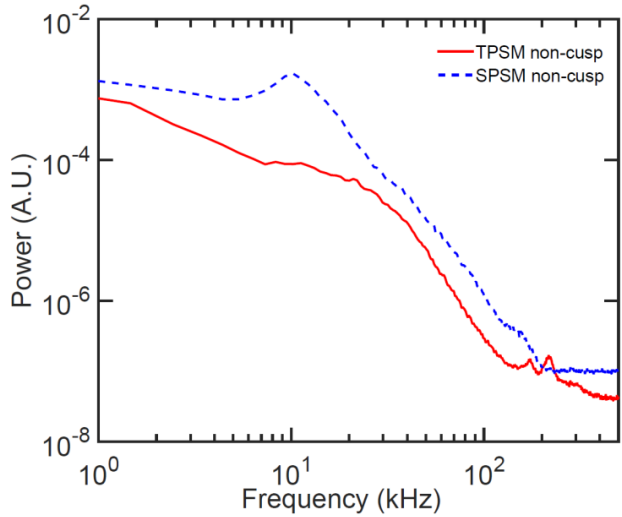

*Figure 9: Auto-power spectrum of density fluctuation at $R = 14$ cm, for pole cusp magnetic field $B_p = 800$ Gauss.*

The auto power spectra of density fluctuation, measured at R=14 cm in non-cusp regime of both configuration, is presented in figure 9, for pole magnetic field $B_P = 800$ G. The figure shows a peak in the range of 10 kHz for SPSM confirming the presence of drift wave while no signature of drift wave is observed in TPSM. The same measurement is performed in the cusp of both configurations as well as for the virtual cusp of TPSM. The signature of the drift wave is not observed in cusp regimes. It can be explained by observing the radial profiles of plasma density and azimuthal profile of ion saturation current that the gradient in the profiles of TPSM is relatively weak to excite any gradient driven fluctuation like the drift wave observed in SPSM.

### ii. Axial Plasma Flow

The net axial plasma flow ($V_d/C_s$) is measured using Mach probe the results are shown in figure 10. It is observed, that the net axial flow in the centre of the device is negligible $< 0.1 C_s$ for both cusp configurations. As we move radially outboard the axial flow increases in SPSM while it remains constant in TPSM. The net axial flow in TPSM is <$0.2 C_s$ throughout the radial extent while for SPSM it increases sharply in the outboard regime. The flow driven by drift wave is found along the z-axis as reported in earlier work [25]. The driver of the mechanism behind this observation is found to be a radial density gradient, in detail, it is discussed earlier. In TPSM the absence of the driver leads to negligible axial flow. It also shows that, when plasma is confined in TPSM, there is no background axial flow.

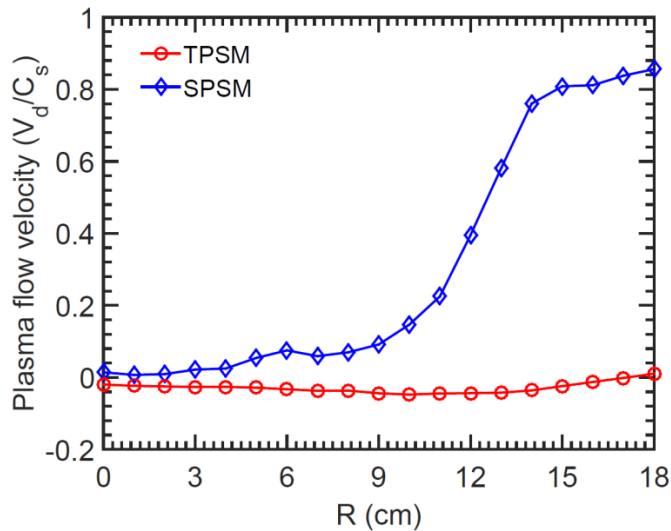

*Figure 10: Radial profile of net axial plasma flow measured using the Mach probe, for pole magnetic field $\boldsymbol{B_p} = \boldsymbol{800}$ G.*

## IV. Summary

The detailed observation of variation of plasma parameters in SPSM and TPSM configurations is carried out. It provides very strong evidence that the magnetic field profile plays a crucial role in the formation of mean plasma profile, plasma fluctuations, and transforming the plasma characteristics. TPSM has larger field free volume and shows more radial uniformity in plasma density with $< 5\%$ variation for the circular area with $R \sim 10$ cm and $\delta I_{isat}/I_{isat} < 1\%$ throughout the plasma. The azimuthal uniformity for a circular area of $\sim 10$ cm radius, absence of drift wave mode, absence of any background plasma flow and fluctuations makes the ideal plasma background to excite any wave mode or perturbation. The examination of plasma parameters suggests that the plasma confined in TPSM is more uniform and quiescent. It is the best configuration for the fundamental plasma study such as excitation of plasma waves, the study of wave-particle interaction, Landau damping etc.

775–8

[18]   Kaw P K 1973 Quasiresonant mode coupling of electron plasma waves *Physics of Fluids* **16** 1967

[19]   Sen Gupta S and Kaw P K 1999 Phase Mixing of Nonlinear Plasma Oscillations in an Arbitrary Mass Ratio Cold Plasma *Physical Review Letters* **82** 1867–70

[20]   Bauer B S, Wong A Y, Decyk V K and Rosenthal G 1992 Experimental observation of superstrong electron plasma waves and wave breaking *Physical Review Letters* **68** 3706–9

[21]   Tajima T and Dawson J M 1979 Laser Electron Accelerator *Physical Review Letters* **43** 267–70

[22]   Tabak M, Hammer J, Glinsky M E, Kruer W L, Wilks S C, Woodworth J, Campbell E M, Perry M D and Mason R J 1994 Ignition and high gain with ultrapowerful lasers* *Physics of Plasmas* **1** 1626–34

[23]   Patel A D, Sharma M, Ramasubramanian N, Ganesh R and Chattopadhyay P K 2018 A new multi-line cusp magnetic field plasma device (MPD) with variable magnetic field *Review of Scientific Instruments* **89** 043510

[24]   Patel A D, Sharma M, Ramasubramanian N, Ghosh J and Chattopadhyay P K 2020 Characterization of argon plasma in a variable multi-pole line cusp magnetic field configuration *Physica Scripta* **95** 035602

[25]   Patel A D, Sharma M, Ganesh R, Ramasubramanian N and Chattopadhyay P K 2018 Experimental observation of drift wave turbulence in an inhomogeneous six-pole cusp magnetic field of MPD *Physics of Plasmas* **25** 112114

[26]   D. Meeker, "Finite element method magnetics," Version 4.2, Users Manual, 2015, http://www.femm.info

[27]   Chung K S 2012 Mach probes *Plasma Sources Science and Technology* **21**
17